\def\cal#1{{\cal #1}}

\def\m@th{\mathsurround=0pt}
\def\n@space{\nulldelimiterspace=0pt \m@th}
\def\biggg#1{{\mbox{$\left#1\vbox to 20.5pt{}\right.\n@space$}}}

\def\beginenum{\begin{enumerate}}
\def\endenum{\end{enumerate}}
\def\bitem{\begin{itemize}}
\def\eitem{\end{itemize}}
\def\bray{\begin{array}}
\def\eray{\end{array}}
\def\begindoc{\begin{document}}
\def\enddoc{\end{document}}
\def\bq{\begin{equation}}
\def\eq{\end{equation}}
\def\bqy{\begin{eqnarray}}
\def\eqy{\end{eqnarray}}
\def\bqyn{\begin{eqnarray*}}
\def\eqyn{\end{eqnarray*}}
\def\bc{\begin{center}}
\def\ec{\end{center}}
\def\bfll{\begin{flushleft}}
\def\efll{\end{flushleft}}
\def\bflr{\begin{flushright}}
\def\eflr{\end{flushright}}
\newcommand{\Avec}{\mbox{\boldmath $A$}}
\newcommand{\Bvec}{\mbox{\boldmath $B$}}
\newcommand{\Evec}{\mbox{\boldmath $E$}}
\newcommand{\Fvec}{\mbox{\boldmath $F$}}
\newcommand{\Gvec}{\mbox{\boldmath $G$}}
\newcommand{\Rvec}{\mbox{\boldmath $R$}}
\newcommand{\Uvec}{\mbox{\boldmath $U$}}
\newcommand{\Vvec}{\mbox{\boldmath $V$}}
\newcommand{\evec}{\mbox{\boldmath $e$}}
\newcommand{\jvec}{\mbox{\boldmath $j$}}
\newcommand{\kvec}{\mbox{\boldmath $k$}}
\newcommand{\nvec}{\mbox{\boldmath $n$}}
\newcommand{\uvec}{\mbox{\boldmath $u$}}
\newcommand{\vvec}{\mbox{\boldmath $v$}}
\newcommand{\wvec}{\mbox{\boldmath $w$}}
\newcommand{\xvec}{\mbox{\boldmath $x$}}
\newcommand{\omegavec}{\mbox{\boldmath $\omega$}}
\newcommand{\Omegavec}{\mbox{\boldmath $\Omega$}}

\documentclass[twocolumn,article]{revtex4}%
\usepackage{amsfonts}
\usepackage{amsmath}
\usepackage{amssymb}
\usepackage{color}
\usepackage{graphicx}%
\begin{document}

\title{Flow Generation via Catastrophic Loss of Equilibrium in
Weakly-Rotating Self-Gravitating Fluids: A Minimal Idealized Model}

%Catastrophic Flow generation in the weakly-rotating,
%self-gravitating neutral fluids }
%surrounding massive compact objects}

\author{L. Gudushauri}
\email{lasha.ghudushauri28@gmail.com}
\affiliation{Department of
Physics, Faculty of Exact and Natural Sciences, Ivane
Javakhishvili Tbilisi State University (TSU), Tbilisi 0179,
Georgia}
\affiliation{E. Kharadze Georgian National Astrophysical Observatory,
Abastumani 0301, Georgia}
\author{N. L. Shatashvili}
\email{nana.shatashvili@tsu.ge}
\affiliation{Department of Physics, Faculty of Exact and Natural Sciences, Ivane
Javakhishvili Tbilisi State University (TSU), Tbilisi 0179, Georgia}
\affiliation{Andronikashvili Institute of Physics, TSU, Tbilisi 0177, Georgia }
\author{G. Shekiladze}
\email{gshekiladze@ufl.edu}
%\affiliation{Andronikashvili Institute of
%Physics, TSU, Tbilisi 0177, Georgia }
\affiliation{Department of
Physics, Faculty of Exact and Natural Sciences, Ivane
Javakhishvili Tbilisi State University (TSU), Tbilisi 0179,
Georgia}
\affiliation{Department of Physics, University of Florida, Ginesville,
Florida, 32611, USA}
\author{S. M. Mahajan}
\email{mahajan@mail.utexas.edu}
\affiliation{Institute for Fusion
Studies, The University of Texas at Austin, Austin,Tx 78712}

\begin{abstract}
This paper explores the catastrophic energy transformations, in particular the
ones leading to the generation of a flow in a weakly rotating
self-gravitating fluid/gas found, for instance, in the vicinity of a massive compact object.
Because of the similarity in the governing equations, the system dynamics is
worked  out exactly in parallel to the methods developed for investigating
catastrophic relaxation in stellar plasmas  \citep{osym,osym2,SSMD}.
In the latter a more ``complex" equilibrium state, on slow changes in the
environment, can lose its equilibrium (catastrophe), and
transform to a less complex state with a very
different energy mix from the original. It is shown that a similar
transformation in the weakly rotating self-gravitating fluid/gas will convert
much of its gravitation energy into kinetic energy in the flow.
Since flows are a perennial ingredient of high-energy astrophysical
systems, the energy transformation processes revealed in present
study, can advance our understanding of a variety of them. Some particularly
relevant examples are: macro-scale flows / structures in galaxies, accretion discs,
and the dynamics and stability of a rotating star / its atmosphere.

\vspace{05.mm}

\noindent {\it Subject Headings}: Vortex Flows - Vortex Dynamics; Alternative
Gravity Theories; Formation and Evolution of Stars \& Galaxies;
Transient and Explosive Astronomical Phenomena; Magnetohydrodynamics
\end{abstract}

%\keywords{Formation and Evolution of Stars \& Galaxies; Vortex Flows - Vortex Dynamics; Alternative
%Gravity Theories; Transient and Explosive Astronomical Phenomena; Magnetohydrodynamics}
%\pacs{04.40.-b, 47.32.−y, 47.65.-d, 52.30.-q, 52.30.Cv, 95.30.Lz, 95.30.Qd, 97.10.Bt, 97.10.Gz, 98.52.-b}
%\pacs{..}
%\startpage{1}
%\endpage{1}
\maketitle

\section{Introduction}

In this paper, we develop a minimal idealized model for catastrophic (when an
equilibrium is suddenly lost) flow generation in a weakly rotating self-gravitating
neutral fluid/gas surrounding astrophysical objects
like extremely massive compact object (Active Galactic Nucleus (AGN),
quasars). Such flows could become the source of relativistic jets
that emerge from these formations; these jets could extend  million parsecs in length.

The accretion disks providing the ``material"" (matter,
energy, and angular momentum) in outflows/jets/winds
(\citep{Begelman,begelman4}), play a crucial role in the  star
formation process - naturally the resulting disk-jet structure,
appropriately, dictates the evolution
of the star \citep{bland,Ustyugova,Dullemond07,jet-photon}.

In order to explore the problem of flow generation, we will employ,
here, a framework similar to the one developed to study catastrophic
(as well as quiescent) phenomena in plasmas constituting the stellar atmospheres
\citep{osym,osym2,msms,SSMD}. The systems have tremendous similarities despite
the fact that the stellar plasmas are controlled, mostly, by electromagnetism
while the dynamics of the weakly rotating self-gravitating neutral
matter in accretion disks is purely gravitational.
This happens because linear but relativistic GR (General Relativity)
is not Newtonian - in the weak-field limit of GR, the rotation of
a self-gravitating fluid can twist the background spacetime surrounding it
(known as frame dragging identified with a magnetic-type gravitational field \citep{Lei,Krishnan},
gravito-magnetic field) this dynamics, governed by the equations derived by taking
the weak-field, slow-velocity limit of Einstein's equation \citep{Chimnoy,sebens2022},
has a structure that has almost one-to one correspondence
with Maxwell equations. Because of this mathematical similarity
in the governing equations, the results are expected
to have much in common. In fact, it was shown
in \cite{SY-DJ} that the systems like the accretion disk-jet outflows, even when
they are vastly different in characteristic parameters (like the Lorentz factor,
Reynolds number, Lundquist number, ionization rate etc.) have a common macroscopic
geometry in addition to a shared mathematical structure.

The  general shared  mathematical structure consists of  what has
been described as Beltrami-Bernoulli (BB) equilibria
\citep{MY,YM,ymois,mmns}; the details of the system  simply change the parameters
without affecting the structure of the system. Since many authors have analyzed BB
equilibria in various settings, we will refer to them at appropriate stage of our calculation.

\section{Description of the theoretical model, model equations}

We will now go directly to the set of equations governing the dynamics of the
weakly rotating self-gravitating matter. Our equations
and initial (equilibrium) methodology is guided by the original paper of \citep{MY}
(recently \citep{Chimnoy} used it for the system
under consideration) while the latter part of the calculation (dealing with
catastrophic loss of equilibrium) is heavily influenced by \citep{osym,osym2}.

When the space-time metric is almost Minkowskian, terms of $O(c^{-4})$ and higher
can be neglected, and the linearized Einstein equations appear almost identical
to Maxwell equations;
\begin{equation}
\nabla \cdot {\bf E}_g = -4\pi G_s \rho \ ,
\label{eq2}
\end{equation}
\begin{equation}
\nabla \cdot \bf{B}_g = 0 \ ,
\label{eq3}
\end{equation}
\begin{equation}
\nabla \times \bf{E}_g = -\frac{1}{c}\frac{\partial}{\partial t}{\bf B}_g \ ,
\label{eq4}
\end{equation}
\begin{equation}
\nabla \times \bf{B}_g= \left(-\frac{4\pi}{c} G_s \rho \bf{v}
+ \frac{1}{c} \frac{\partial}{\partial t}{\bf E}_g\right) \ .
\label{eq5}
\end{equation}
The total gravitational force - ``Lorentz force'' acting
on a fluid particle is:
\[
\textbf{f}_g = \rho \left( \bf{E}_g + \frac{1 }{c}\bf{v}\times \bf{B}_g \right) \ .
\label{eq16}
\]
Also, the equation of motion governing the self-gravitation charge less fluid
\citep{sebens2022,heaviside}:
\begin{equation}
\rho \left( \frac{\partial \bf{v}}{\partial t} + (\bf{v} \cdot \nabla )\bf{v} \right)
= - \nabla p - \rho \left( \bf{E}_g + \frac{1}{c}\bf{v} \times \bf{B}_g \right) \ ,
\label{eq6}
\end{equation}
is simply a replica of the equivalent ``Lorentz force'' equation of a
negatively charged fluid \cite{sebens2022}.
The gravitating fluid obeys the Continuity Equation (holds everywhere):
\begin{equation}
\frac{\partial \rho}{\partial t} + \nabla \cdot (\rho \bf{v}) = 0 \ .
\label{17}
\end{equation}
In the preceding equations,  ${\bf v}$  is the fluid velocity,
$\rho$ is the mass density of neutral matter, $p $
is local fluid-pressure, $\bf{E}_g = -\nabla \phi_g -
\frac{1}{c}\frac{\partial \textbf{A}_g}{\partial t}$
is the local gravitational field (so called ``gravito-electric'' field)
with the corresponding local gravitational potential $\phi_g$ and $\bf{B}_g
= \nabla \times \textbf{A}_g $ is the ``gravito-magnetic'' field with the
corresponding $ \textbf{A}_g$  ``vector'' potential. Hence, we are taking
into account the self-gravity of the fluid/gas, with $G_s$ being
the gravitational constant characterizing the fluid/gas.
Equations (\ref{eq2})-(\ref{17}) (reflecting the attractive
nature of gravity) will  serve as the foundations
for this work; following the methodology of \citep{osym,osym2,ymois,msms},
we will now formulate the gravitational equivalent of catastrophic
flow generation. We will also compare and contrast the behavior of
flows generated by weakly rotating self-gravitating
and electromagnetic systems.
Notice, that in present paper we study the equilibrium configurations of
a weakly rotating self-gravitating neutral fluid in the weak field limit;
in the PPN coordinate frame, in the weak field slow motion limit, Equations
(\ref{eq2})-(\ref{17}) with the corresponding scalar and vector potentials
of ``Gravitoelectromagnetic field"" (GEM) look like Maxwell equations
(\citep{thorne1986,thorne1988,Braginsky,manfredi,manfredi2013,sebens2022,heaviside}
and references therein). We understand that for the stationary background
space-time, the time-dependent term in (\ref{eq4}) is absent.
In our equilibrium analysis, this term does not appear anyway (it is written
just to display the similarity to Maxwell equations). However, in the time
dependent analysis we plan to carry out in our next paper, this term
though small is likely to play a significant role. Even though the effects
of GEM are small, their study is important for more complex gravitational
phenomena on different scales (in \cite{Chinmoy2} this issue is well discussed).
In fact, in our next paper, we will emphasize the new physics that originates
in GEM. In the present equilibrium analysis, it does not play a role, and
the “gravito-electric limit” is sufficient.

\section{Quasi-equilibrium analysis - explosive event leading to
catastrophic formation of flow}

In analogy with \citep{osym,osym2}, we expect  two distinct scenarios
for explosive/eruptive events for the weakly-rotating self-gravitating system:
(i) when a slowly evolving structure finds itself in a state
of no equilibrium, and (ii) when the process of creating a long-lived
structure is prematurely aborted; the flow shringks/distorts
the structure, which suddenly shines and/or released energy
or ejects particles. The latter mechanism requires a detailed
time-dependent treatment and is not the subject matter of
this study. The former semi-equilibrium, collisionless
gravito-magnetofluid treatment will be pursued here.
For simplicity, our analysis is limited to constant density structures.
Though somewhat  drastic, this  step will help us capture the
underlying physics in a much simpler analytically tractable
calculation. Including density variation, though, is straightforward
\citep{mmns,mnsy,BS-flow}.

\bigskip

\subsection{Beltrami-Bernoulli Equilibrium States}

We begin by introducing some definitions: \ $\textbf{P}_g = \bf{v}
- \frac{\bf{A}_g}{c}$ \ is the generalized momentum,  \
$\Phi = \frac{p}{\rho} + \frac{v^2}{2} - \phi_g$ \ is the generalized
pressure, and
\begin{equation}
\bf{\Omega} = -\frac{\bf{B}_g}{c} + \nabla \times \bf{v} \ .
\label{eq22}
\end{equation}
is the Generalized Vorticity (analog of the Generalized Vorticity
of negatively charged fluid).

Then, using  Equations (\ref{eq2}) -(\ref{eq5}), we can derive a stationary
Double Beltrami (DB) equilibrium - a Beltrami equilibrium results from
the alignment of an effective velocity along its (generalized) vorticity:
\ $\bf{\Omega} = - |b| \rho \bf{v}$, the constant of proportionality $b$
is a parameter of the system.  The equilibrium does, simultaneously,
satisfy the Bernoulli condition - the vanishing of gradient forces
($\nabla \Phi = 0$) and the time-independent Continuity Equation
$\nabla (\rho \bf{v}) = 0$. The principal result of the procedure,
however, is the emergence of the promised Double Beltrami (DB) equation
\begin{equation}
\nabla \times \nabla \times \bf{v} + |b|\nabla \times (\rho \bf{v})
+ \frac{4\pi G_s \rho}{c}\bf{v} = 0 \ ,
\label{eq28}
\end{equation}
the preceding equation, in this context, was first derived by
\cite{Chimnoy} to describe the relaxation of
self-gravitating weakly rotating astrophysical flows.
Notice, that all the terms in (\ref{eq28}) are with positive sign that
is because of the minus sign of current in ``Amp\'ere's law''
(see the r.h.s. of the Eq. (\ref{eq5})).
DB equation and its solutions were studied in detail in the context of
Hall MHD (see e.g. \citep{mmns,ymois,mnsy,msms}) both for Astrophysical
and laboratory plasmas
\footnote{$^1$ We remind the reader that for multi-fluid plasmas
each Beltrami condition is an alignment of a vorticity and
a corresponding flow velocity. Our system consists of only one neutral fluid, hence,
we have only one velocity ${\bf v}$ of this fluid locally, at the same time,
we have a so called gravito-magnetic field ($\bf{B}_g \neq 0$ \ due
to self-gravity taken into account, $\kappa \neq 0 $ )  and, as shown above,
the flow-vorticity is generalized (see (\ref{eq22})); also we have only one
Beltrami paramater $b$. Then, the
derived equation (\ref{eq33}) is of higher order (DB-like);  without this
effect (leading to $\kappa \to 0$) we would get the Single Beltrami (SB) equation
for the flow (see the details of derivation in \cite{Chimnoy}).
We can, however, mathematically analyze our self-gravitating flow as a DB system -
flow generalized by self-gravity - and follow the methodology developed
for the exploxive/erruptive events.}.

By assuming $\rho = \rho_0 \approx const$, and introducing normalized variables
\[
\textbf{r}^{'} \to \lambda_j^{-1}\,\textbf{r} \ , \qquad  \bf{v}^{'} \to \frac{\bf{v}}{c} \ ,
\qquad  \bf{B}_g^{'} \to \frac{\lambda_j}{c^2}\, \bf{B}_g ,
\]
\begin{equation}
p^{'} \to \frac{p}{\rho_0 c^2} \ , \qquad |b|\rho_0 \lambda_j \equiv b^{'} > 0
\label{eq31}
\end{equation}
in terms of the characteristic parameters of the system: $\omega_j =
\sqrt{4\pi G_s \rho_0}$ (Jean's frequency), $\lambda_j = c_{s0}/\omega_{j}$,
(Jean's length) (where $c_{s0}$ is the speed of sound), and
\[
\kappa = \frac{\omega_j^2 \lambda_j^2}{c^2} = \frac{c_{s0}^2}{c^2} \ll 1 \ ,
\]
Equation (\ref{eq28}) is written in a dimensionless
form ($b > 0$ and $\kappa > 0$ by definition) as follows (primes are omitted):
\begin{equation}
\nabla \times \nabla \times \bf{v}+ b\nabla \times \bf{v} + \kappa \bf{v} = 0 \ .
\label{eq33}
\end{equation}
Comparing this equation with the DB equation used for the two-fluid plasma \cite{osym}
we find a sign difference in the last terms; also the sign of the 2nd term
is now fixed. For the neutral fluid both terms are with positive signs.
We will see the consequences of this effect below.

Equation (\ref{eq33}) can be formally written as  ($curl = \nabla \times $)
\begin{equation}
(curl - \lambda_+)(curl + \lambda_-)\bf{v} = 0 \ ,
\label{eq41}
\end{equation}
where, $\lambda_+$ and $\lambda_-$ are the solutions of quadratic equation:
\begin{equation}
\lambda^2 + b\lambda + \kappa = 0 \ ,
\label{eq42}
\end{equation}
yielding ( $b > 0$)
\begin{equation}
\lambda_{\pm} = \frac{1}{2}(-b \pm \sqrt{b^2 - 4\kappa}) \ .
\label{eq43}
\end{equation}

The general solution of the DB equation is a linear combination of single
Beltrami fields $(\bf{G}_{\pm})$, which satisfy the following condition:
$\nabla \times \bf{G}_{\pm} = \lambda_{\pm} \bf{G}_{\pm}$
\cite{MYPOP,ymois}. The velocity  {\bf{v}}, then, may be expressed as a linear superposition
\begin{equation}
\bf{v} = C_+ \bf{G}_+ + C_- \bf{G}_- \ .
\label{eq44}
\end{equation}
Using (\ref{eq44}) and the expression for vorticity $\bf{\Omega} = -b \bf{v}$,
the Generalized vorticity
\begin{equation}
\bf{B}_g = b\bf{v} + \nabla \times \bf{v} \
\label{eq45}
\end{equation}
becomes:
\begin{equation}
\bf{B}_g = (b + \lambda_+)C_+\bf{G}_++(b + \lambda_-)C_-\bf{G}_- \ .
\label{eq46}
\end{equation}

The non-zero coefficient $\kappa$ contains the Jean's length
$\lambda_j$ that is not zero due to the inclusion of fluid
self-gravity. The relation between $\kappa $ and $b$ will determine
the characteristic scales (could be highly separated) of the system.

If  $\frac{\kappa}{b^2} \ll 1$, \ the roots $\lambda_{\pm}$ will
be vastly separated (system exhibits the macro- and micro-scale structures):
\begin{equation}
\lambda_+ = \frac{1}{2}(-b + \sqrt{b^2 - 4\kappa}) \approx -\frac{\kappa}{b}
< 0 \ ,
\label{eq48}
\end{equation}
\begin{equation}
\lambda_- = \frac{1}{2}(-b - \sqrt{b^2 - 4\kappa}) \approx -b  < 0\ .
\label{eq49}
\end{equation}
implying
\begin{equation}
|\lambda_-| \gg |\lambda_+|;
\label{eq50}
\end{equation}
admitting  macro- (inverse of $|\lambda_+|$) and micro-
(inverse of $|\lambda_-|$) structures.

\subsection{Conservation Laws}

We consider adiabatic processes when the parameters change quite slowly
assuming that a given structure at every stage can be described by
the DB equilibrium solutions \citep{osym,osym2}.
For a system governed by DB equation, there are three
conserved quantities: the total energy and two helicities.
The general DB field is characterized by four parameters
- with its eigenvalues $(\lambda_{\pm})$ and amplitudes $(C_{\pm})$
\cite{Stein}.
The helicity is generally defined as follows (with ${\bf \Omega }$ being the
vorticity):
\begin{equation}
h = \frac{1}{2}\,\int\left(curl^{-1} {\bf \Omega} \right)\cdot {\bf \Omega}
\, d\textbf{r} \ .
\label{eq51}
\end{equation}
For the {\it Total Energy} of our weakly-rotating, self-gravitating fluid we have
following \citep{thorne1988,manfredi,sebens2022,Chimnoy}:
\begin{equation}
E = \frac{1}{2}\int\left(v^2 - \frac{B_{g}^{2}}{\kappa}\right) \ .
\label{eq55}
\end{equation}
Notice, that unlike the conventional plasma magnetic field energy, the gravito-magnetic
field energy is negative reflecting  the attractive
character of gravity \citep{thorne1988,sebens2022}. The conventional fluid helicity
\begin{equation}
h_{min} = \frac{1}{2}\,\int\bf{v} \cdot(\nabla \times {\bf v})d\textbf{r}
\label{eq56}
\end{equation}
is also changed to
\begin{equation}
h_{gen} = - \frac{b}{2}\, \int \left(v^2 - \frac{B_{g}^{2}}{\kappa}\right)d\textbf{r}
= - b E \
\label{eq61}
\end{equation}
for the self-gravitating fluid.

For simplicity, our domain is a  Cartesian cube of length L (similar to \citep{osym,osym2}).
We model the Beltrami field $G_{\pm}$ by  the simple two-dimensional
ABC field (\cite{Arnold}),
\begin{equation}
\begin{split}
\textbf{G}_{\pm} = \textbf{g}_{x \pm}
\begin{pmatrix}
0 \\
sin(\lambda_{\pm} x) \\
cos(\lambda_{\pm} x)
\end{pmatrix}
+ \textbf{g}_{y \pm}
\begin{pmatrix}
cos(\lambda_{\pm} y )\\
0 \\
sin(\lambda_{\pm} y )
\end{pmatrix}  \quad ,
\end{split}
\label{eq62}
\end{equation}
with the normalization $|\textbf{g}_{x \pm}|^2 + |\textbf{g}_{y \pm}|^2 = 1$.
Periodic boundaries condition  \ \ $L = n_+ \frac{2\pi}{\lambda_+}
= n_- \frac{2\pi}{\lambda_-}$ define the intergers
with $n_{\pm}$ ;  in addition,
$\int (\bf{G}_+ \cdot \bf{G}_-) d\textbf{r} = 0 $,
\ \ $\int \bf{G}_{\pm}^2 d\textbf{r} = L^2$.

In terms of the parameters of the DB solutions, we may write the minimal helicity
(fluid without self-gravity) and the generalized (fluid with self-gravity) helicity:

\begin{equation}
h_{min} =
%\frac{1}{2} \int\bf{v} \cdot( \nabla \times \bf{v}) d\bf{r} =
\frac{L^2}{2}\,(\lambda_+ C_+^2 + \lambda_- C_-^2) \ ,
\label{eq63}
\end{equation}
\[
h_{gen} = h_{min} - L^2\left( (b + \lambda_+)^2 C_+^2 + (b + \lambda_-)^2 C_-^2 \right) +
\]
\begin{equation}
+ \frac{L^2}{2}\left( (b + \lambda_+)^2 \frac{C_+^2}{\lambda_+} + (b + \lambda_-)^2
\frac{C_-^2}{\lambda_-}  \right),
\label{eqGen}
\end{equation}
and, then,
the amplitudes $C_+^2$ and $C_-^2$ are now expressible in terms of helicities:
\begin{equation}
C_{\pm}^2 = \frac{2h_{min}}{L^2 \lambda_{\pm}} - \frac{\lambda_{\mp}}{\lambda_{\pm}}C_{\mp}^2 \ .
\label{eq64}
\end{equation}

Using (\ref{eq46}) and (\ref{eq64}) in (\ref{eq55}) we obtain:
\begin{equation}
\begin{split}
E = \frac{L^2}{2}\left(1 - \frac{(b+\lambda_+)^2}{\kappa}\right)C_+^2 \\
 - \frac{L^2}{2}\left(1 - \frac{(b+\lambda_-)^2}{\kappa}\right)
\frac{\lambda_+}{\lambda_-}C_+^2   \\
+ \left(1 - \frac{(b + \lambda_-)^2}{\kappa}\right
)\frac{h_{min}}{\lambda_-} \ ,
\end{split}
\label{eq65}
\end{equation}
from which, using (\ref{eq64}) and following relations
\[
\lambda_+ + \lambda_- = - b \ , \qquad \lambda_+  \lambda_- = \kappa \ ,
\]
we derive:
\begin{equation}
\begin{split}
C_+^2 = \frac{2\lambda_-}{L^2}\, Q^{-1}\left( E - \frac{h_{min}}{\lambda_-}
\left(1 - \frac{\lambda_+^2}{\kappa}\right)\right) \ ,
\end{split}
\label{eq66}
\end{equation}
\begin{equation}
\begin{split}
C_-^2 = -\frac{2\lambda_+}{L^2}\, Q^{-1}\left( E - \frac{h_{min}}{\lambda_+}
\left(1 - \frac{\lambda_-^2}{\kappa}\right)\right) \ ,
\end{split}
\label{eq67}
\end{equation}
where
\[
Q = \lambda_- \left( 1 - \frac{\lambda_-^2}{\kappa}\right) -
\lambda_+ \left(1 - \frac{\lambda_+^2}{\kappa}\right) =
\]
\begin{equation}
= (\lambda_- - \lambda_+)\left(2 - \frac{b^2}{\kappa}\right) \ .
\label{eqQ}
\end{equation}
Notice, that when
$\frac{\kappa}{b^2} \ll 1$, $\lambda_-<0$ and (\ref{eq50}) is satisfied,
$Q>0$; also $Q^{-1}$ diverges at the coalescence of
the roots (when $\lambda_+ = \lambda_-$). For an acceptable
equilibrium the amplitudes $C_{\pm}$ and the micro-length
(corresponding to one of the $\lambda_{\pm} $) must remain real
as the other goes over real values. Therefore, $C_{\pm}^2$ must
remain positive. When we deal with the equilibria with vastly
separated scales, slow changes in the parameters can cause the loss of equilibrium
only by making one of the Beltrami components to vanish (either of $C_{\pm}^2$
becomes zero starting from positive values) for real $\lambda_{\pm} $.
The other possible route for losing equilibrium when two solutions
become the same ($\lambda_{\pm} $ coalesce), is not likely to happen via adiabatic
changes.

\section{Analysis of catastrophic loss of equilibrium}

For $\kappa \ll b^2$ the roots are real and vastly separated; then the
only catastrophe allowed in this setup is when the amplitude corresponding
to any state/structure approaches zero.

In what follows, $\lambda (\ \equiv \lambda_+)$ denotes  the macro, and
$\mu (\ \equiv \lambda_- )$ the micro-scale; these are related to the
basic parameters of the equilibrium via
\begin{equation}
\begin{split}
\lambda + \mu = - b \ ; \qquad \lambda \mu = \kappa \ .
\end{split}
\label{eq68}
\end{equation}
The catastrophic loss of equilibrium (the higher order
relaxed state is reduced to lower order one) will occur at
either  $C_{\mu}^{2} = 0$ - the disappearance
of the micro-scale constituent of the DB field, or when
$C_{\lambda}^{2} = 0$ implying the disappearance of the macro-scale
constituent of DB field - the DB state, then, must relax to a Single Beltrami
state. When either of them goes to zero, the energy mix of the State must change.
The change could lead to state richer in gravitational (fluid) energy - the
flow generation must necessarily mean that the final state must have
larger fluid kinetic energy than the original one. The catastrophe condition
is met at the critical value of control parameter when (see
(\ref{eq67}) for the former scenario):
\begin{equation}
E - \frac{h_{min}}{ \lambda_{crit}}\left(1 - \frac{\mu^2}{\kappa}\right) = 0 \ ,
\label{eq69}
\end{equation}
which, using (\ref{eq68}), reduces to the following equation
in terms of the  energy and  helicities characterizing the
DB state:
\[
\frac{h_{min}}{\kappa}\lambda_{crit}^2 \ + \ \left( E + 2 b \frac{h_{min}}{\kappa} \right)
\lambda_{crit} \ +
\]
\begin{equation}
+ \ h_{min}\left( \frac{b^2}{\kappa} - 1 \right) \ = \ 0 \ .
\label{eq70}
\end{equation}
Solving the above equation yields an expression for the critical
value of control parameter ($\lambda = \lambda_{crit}$):
\begin{equation}
\begin{split}
\lambda_{crit} = \frac{-F + \sqrt{D}}{2K} \ .
\end{split}
\label{eq71}
\end{equation}
where
\[D = F^2 - 4KM , \qquad K = \frac{h_{min}}{\kappa} ,
\]
\[
F = \left( E + 2b\frac{h_{min}}{\kappa} \right) ,  \qquad
M = h_{min}\left( \frac{b^2}{\kappa} - 1 \right) \ ,
\]
which must be satisfied together with the equations (\ref{eqGen})
and (\ref{eq65}).

The unfolding of the process of the loss of equilibrium for the
self-gravitating weakly rotating fluid follows the same path as
described in \cite{osym2}. Starting from a positive
$C_{\mu (\lambda)}^2$, the ambience must change to push
it past the critical point $C_{\mu (\lambda)}^2=0$.
Negative  $C_{\mu (\lambda)}^2$ implies an imaginary
$C_{\mu (\lambda)}$ and the loss of original
equilibrium.

Notice, that due to the attractive nature of gravity, the current
system allows a negative (total) energy state (different from the
electromagnetic plasma state) for weak background flows with
strong self-gravity; we also emphasize that due to (\ref{eq48},\ref{eq49})
and (\ref{eq63}) both $\lambda $ and $\mu $ are negative as well $h_{min} <0$
with $C_{\mu(\lambda)}^2 >0$.

Qualitative analytic results were strengthened by numerical experiments.
For this paper, meant to theoretically demonstrate a new physical
effect, our choice of $\kappa $ was dictated by considerations of highlighting
the importance of self-gravity. This may not correspond to values realistic
for neutral weakly rotating astrophysical fluids. We do plan to survey
a range of $\kappa $ values in a later publication. The reader may note
that $\kappa = 0.33$ for the relativistic gas.

In FIG.1 and FIG.2,  we illustrate the behavior of  a special  equilibrium with
catastrophe-prone parameters: $E = -3 , \ h_{min} = - 0.5 ,
\ \kappa = 0.04, \ b = 0.5 , \ \lambda_{crit} = - 0.298$.
Here $L^2/2$ is normalized out and it was set that $C_{\lambda ,\mu } > 0$.
In FIG.3 and FIG.4 we present the results for
the parameters: $E = -13 , \ h_{min} = - 0.5 , \ \kappa = 0.2, \ b = 5  ,
\ \lambda_{crit} = -3.14$. In both cases the macro scale constituent completely
vanishes while the micro-scale structures strengthen (see Figures 1,3). One can see
that it is the micro-scale flow energy that slowly increases
feeding the local turbulence (see Figures 2,4) extremely important for
accretion disk local viscosity and disk-jet structure formation
\citep{Shakura1973,yso,jet-photon}.

%%%%%%%%%%%%% FIG1 %%%%%%%%%%%%%%%%%%%%%%%%%%

\begin{figure}
\begin{center}
\includegraphics[scale=0.55,angle=0]{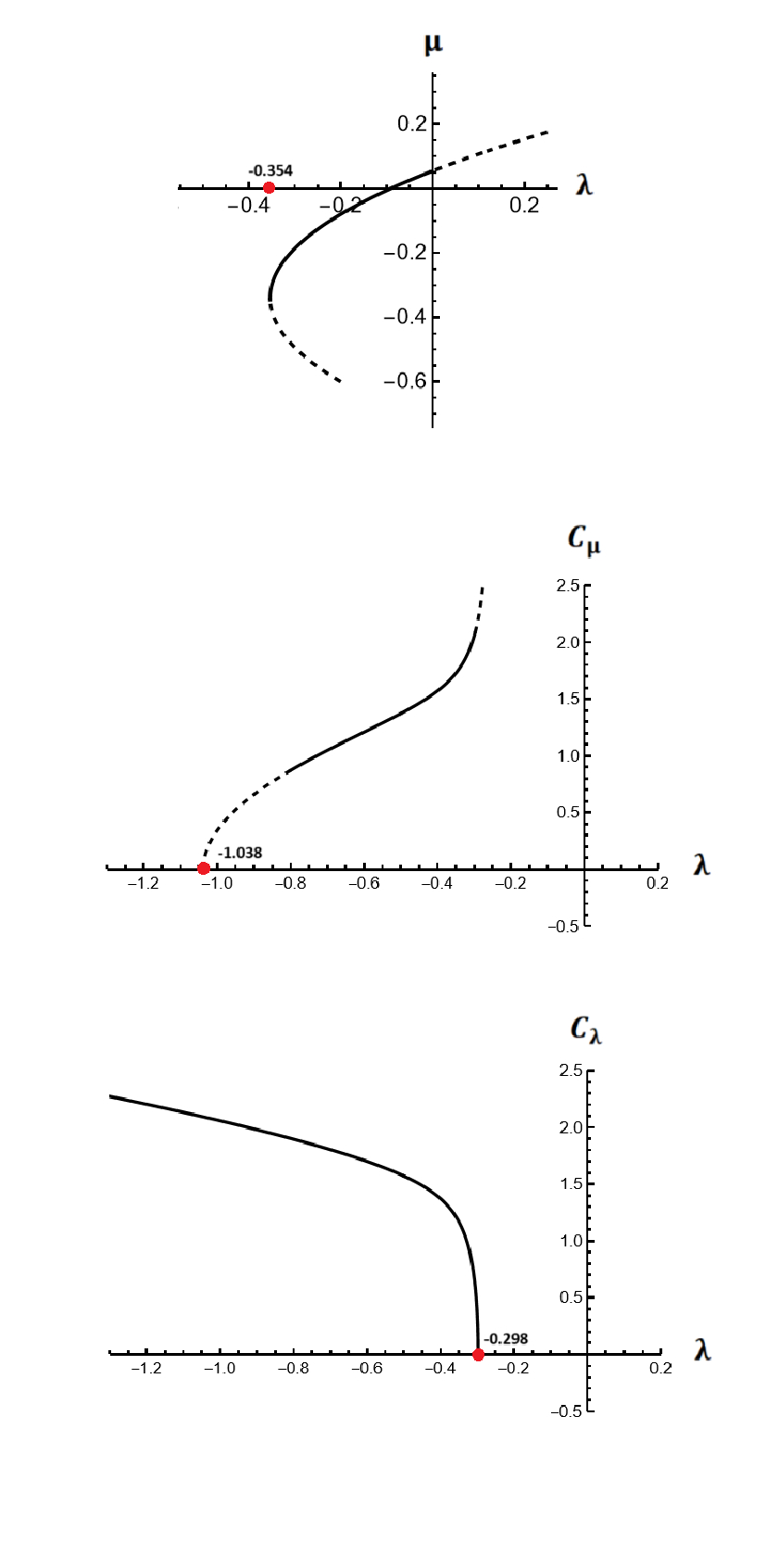}
\caption{Plots for micro-scale $\mu $ and $C_{\lambda,\mu}$ vs macro-scale
$\lambda $ for catastrophic rearrangement of the original state; $E = -3 > E_{crit},
\ h_{min} = - 0.5 , \ \kappa = 0.04 , \ b = 0.5$. \ The critical point
$\lambda_{crit} = -0.298$ at which the transition happens can be observed
on the plot of $C_{\mu}$; the macro-scale constituent is fully vanished.}
\label{Fig.1}
\end{center}
\end{figure}

%%%%%%%%%%%%% FIG2 %%%%%%%%%%%%%%%%%%%%%%%%%%

\begin{figure}
\begin{center}
\includegraphics[scale=0.38,angle=0]{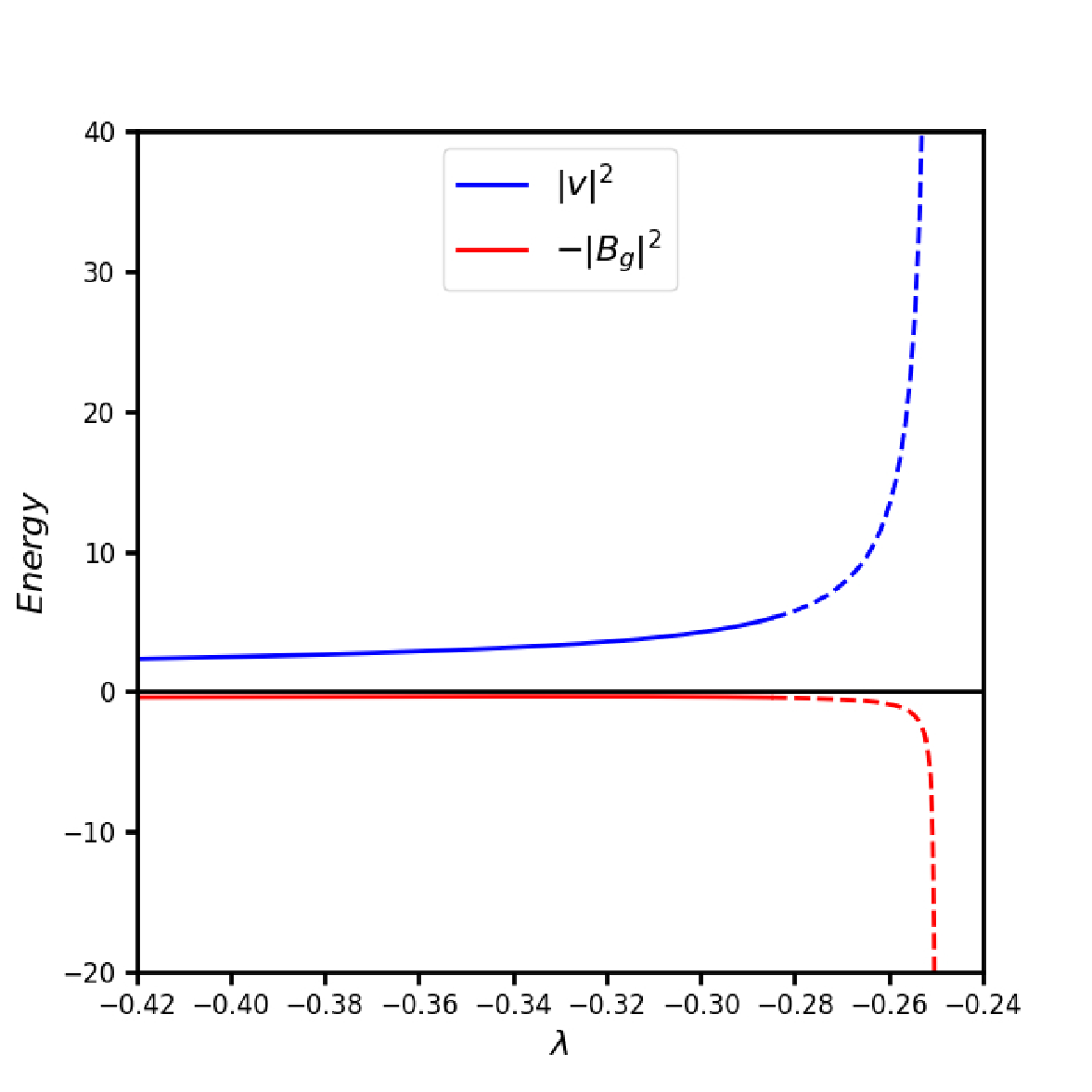}
\caption{Plots for flow and gravito-magnetic energies %multiplied by $\kappa 10^2$
vs $\lambda $ for catastrophe-prone parameters (see Fig1) -
the micro-scale/turbulent flow generation is clearly seen.}
\label{Fig.2a}
\end{center}
\end{figure}

Using the value of $\lambda_{crit}$, we may confirm that, at the critical
point, the coefficient $C_{\lambda}$ which measures
the strength of the macro-scale fields, identically vanishes,
and the equilibrium changes from DB to a SB
state labeled by $\lambda = \lambda_{crit}$, i.e., ${\bf v} =
C_{\mu}\,G_{\mu}$ or $\nabla \times {\bf v} = \mu {\bf v}$ .
The amplitude of the micro-scale component (the only one remaining)
turns out to be $C_{\mu}^2 = \lambda_{crit}^{-1}\,h_{min}$.
The transition leads to a gravito-magnetically more relaxed state with
the gravito-magnetic energy reaching its minimum (taking into account
that it has a ``minus'' sign in total energy) with an appropriate gain in the
flow kinetic energy (turbulent). At the transition, the
kinetic energy is significantly greater than the gravito-magnetic
energy with the ratio $-B_g^2 / v^2 \sim -(b+\mu)^2$ (with $b <1 $;
$|\mu | <1$; see  Figures 2,4).
Starting from vastly separated
scales (at $\kappa \ll b^2$), the initial gravito-magnetic energy is
transferred mainly to flow energy of the microscopic scale.

In both cases, the inverse macro-scale $|\lambda|$  decreased
as we reached the catastrophic transition to the SB state (starting
from the DB state). In the complementary case when the initial conditions are
favorable to a catastrophic change with increasing  $|\lambda|$  (see Fig.5
following the variation of this inverse-scale from right to left, starting from $0$),
it is the macro-scale energy that increases, and the
micro-scale energy practically vanishes;
parameters are the same as for the Figures 3, 4 but the
range of control parameter is different with $\lambda_{crit} = -1.86$.
Specifically interesting is that for such scenarios it is the macro-scale
flow energy that increases and the macro-scale gravito-magnetic energy
decreases with increasing $|\lambda|$ ( $-|{\bf B}_g|^2$, predominantly
macro-scale, is following the flow energy variation in the
opposite direction) - a physically reliable transition.
Note, that in such scenario the macro-scale
structure shrinks and is not destroyed.

%%%%%%%%%%%%%%%%%%%%%%  FIG.3  %%%%%%%%%%%%%%%%%%%%%%%%

\begin{figure}
\begin{center}
\includegraphics[scale=0.55,angle=0]{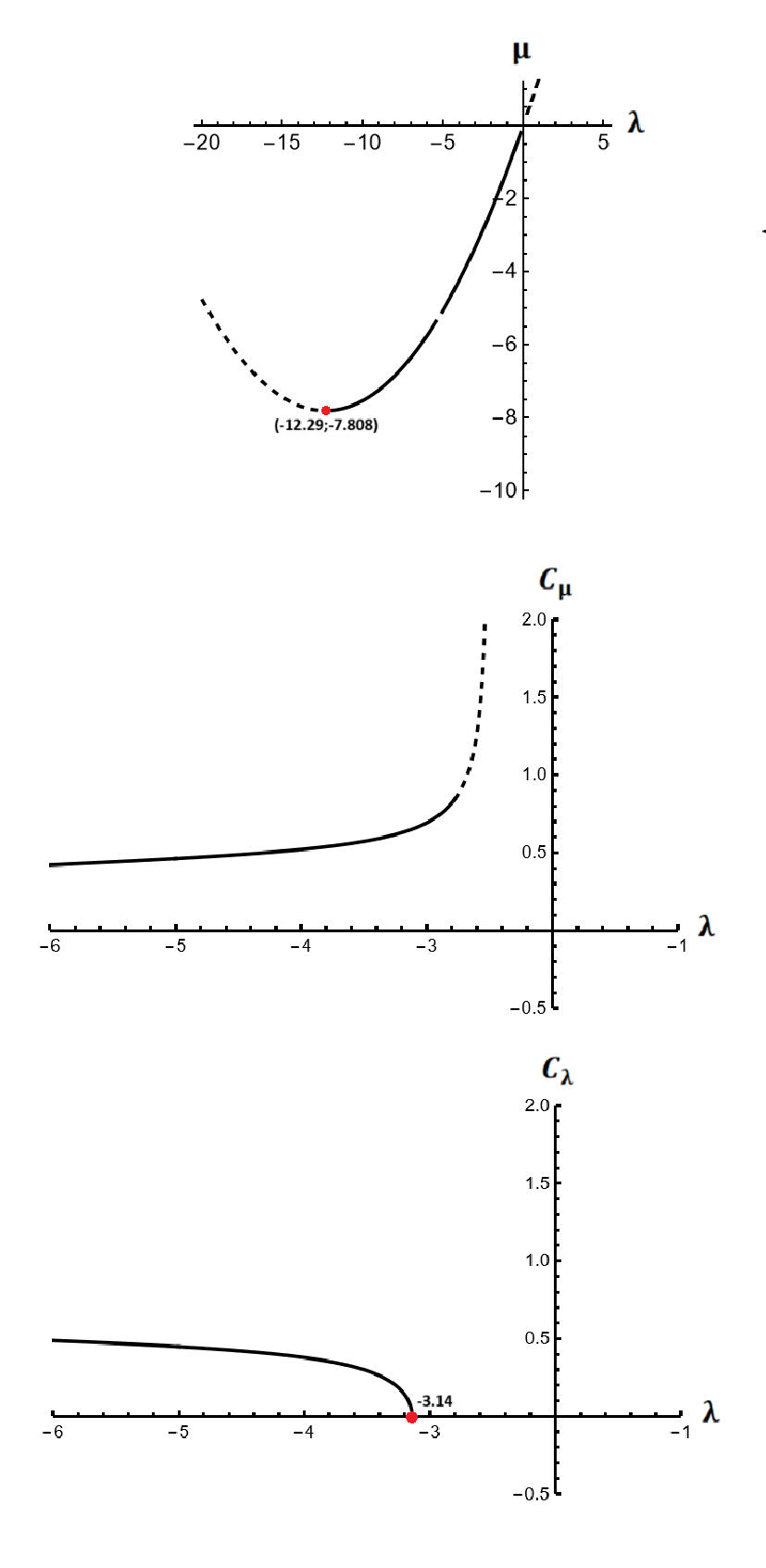}
\caption{Plots for micro-scale $\mu $ and $C_{\lambda,\mu}$ vs macro-scale
$\lambda $ for catastrophic rearrangement of the original state; $E = -13 ,
\ h_{min} = - 0.5 , \ \kappa = 0.2 , \ b = 5$. \ The point
$\lambda_{crit}  = -3.14$ at which the transition to turbulent state happens can
be observed on the plot for $C_{\lambda }$.}
\label{Fig.3}
\end{center}
\end{figure}

%%%%%%%%%%%%%%%%%%%%%%  FIG.4  %%%%%%%%%%%%%%%%%%%%%%%%%%%

\begin{figure}
\begin{center}
\includegraphics[scale=0.38,angle=0]{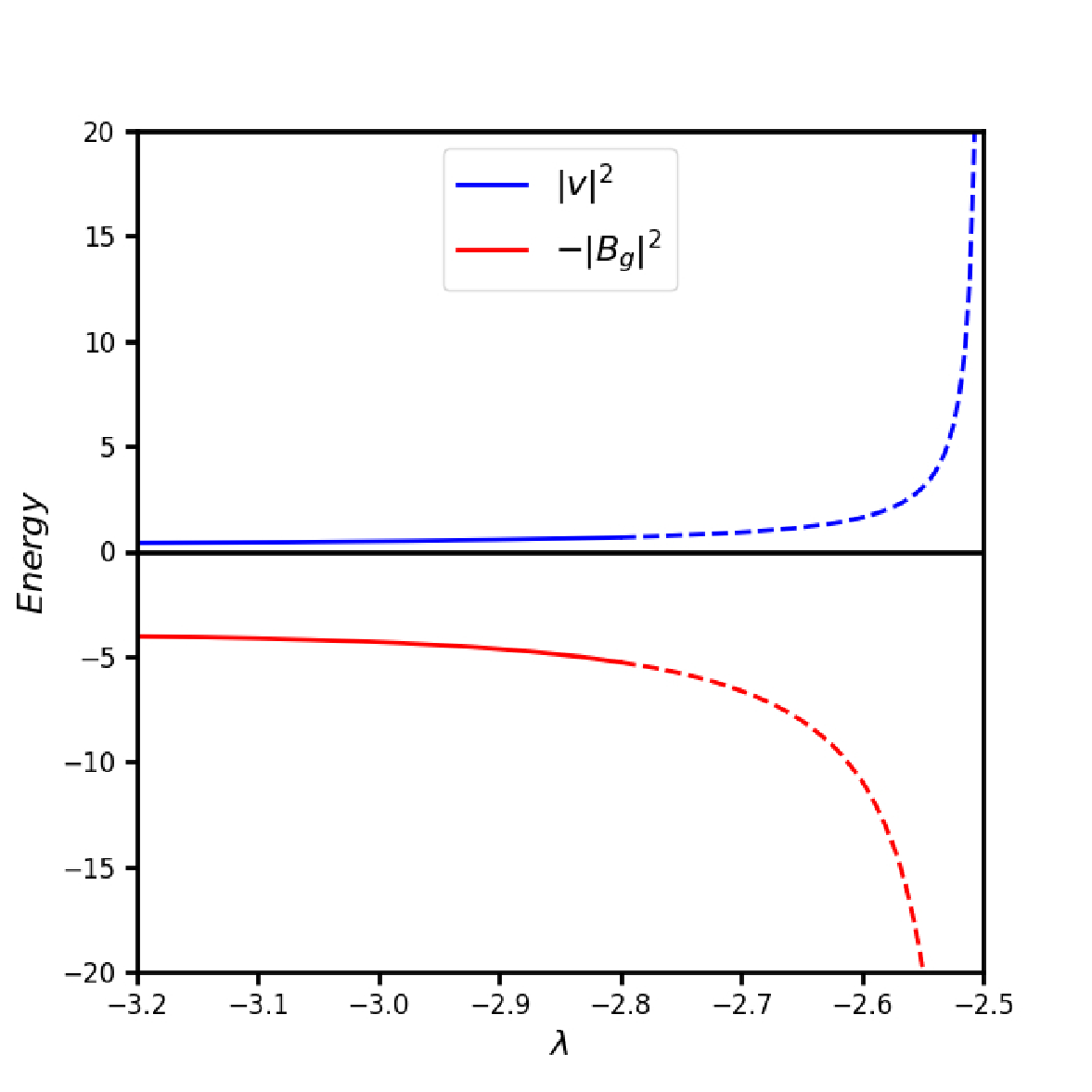}
\caption{Plots for flow and gravito-magnetic energies %multiplied by $\kappa 10^2$
vs $\lambda $ for the scenario leading to final turbulent state (see parameters in Fig.3)
- flow energy is totally micro-scale.}
\label{Fig.4}
\end{center}
\end{figure}

From these figures it is clear that physical parameters of
the system are precisely defined at the critical point.
Also, the representation of a changing structure by a DB
field persists and the length scales remain well separated
since we assume that changes are slow and transport
processes can be ignored (see Fig-s. 1, 3, 5).

%%%%%%%%%%%%%%%%%%%%%%  FIG.5  %%%%%%%%%%%%%%%%%%%%%%%%

\begin{figure}
\begin{center}
\includegraphics[scale=0.55,angle=0]{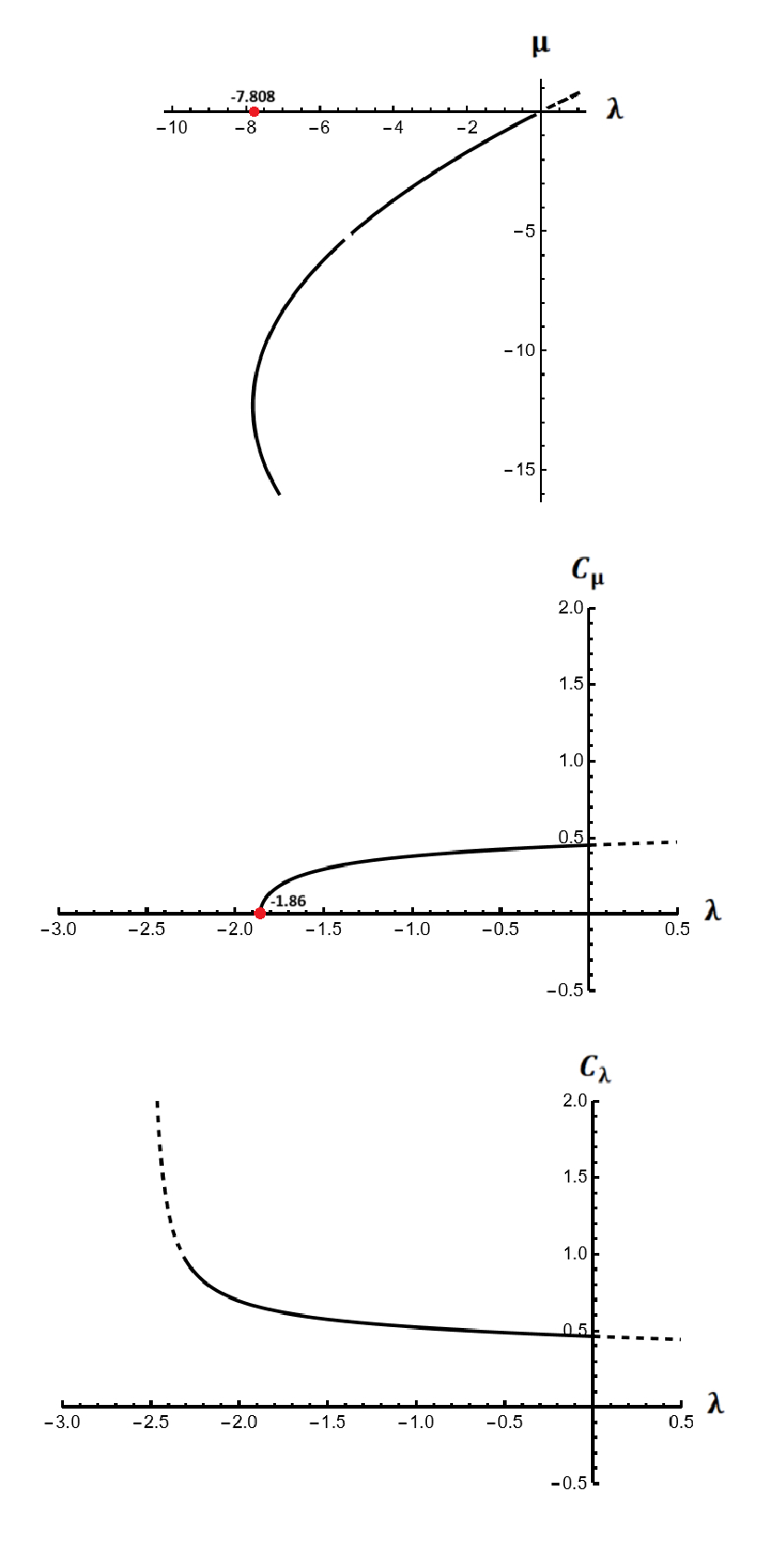}
\caption{Plots for micro-scale $\mu $ and $C_{\lambda,\mu}$ vs macro-scale
$\lambda $ for catastrophic rearrangement of the original state; $E = -13 ,
\ h_{min} = - 0.5 , \ \kappa = 0.2 , \ b = 5$. \ The point
$\lambda_{crit}  = -1.86$ at which the transition to macro-scale structure
dominant state happens can be observed on the plot for $C_{\lambda }$
(when we follow the control parameter $\lambda $ variation from right to
left - following the increase of inverse-length scale $|\lambda |$).}
\label{Fig.5}
\end{center}
\end{figure}

%%%%%%%%%%%%%%%%%%%%%%  FIG.6  %%%%%%%%%%%%%%%%%%%%%%%%%%%

\begin{figure}
\begin{center}
\includegraphics[scale=0.38,angle=0]{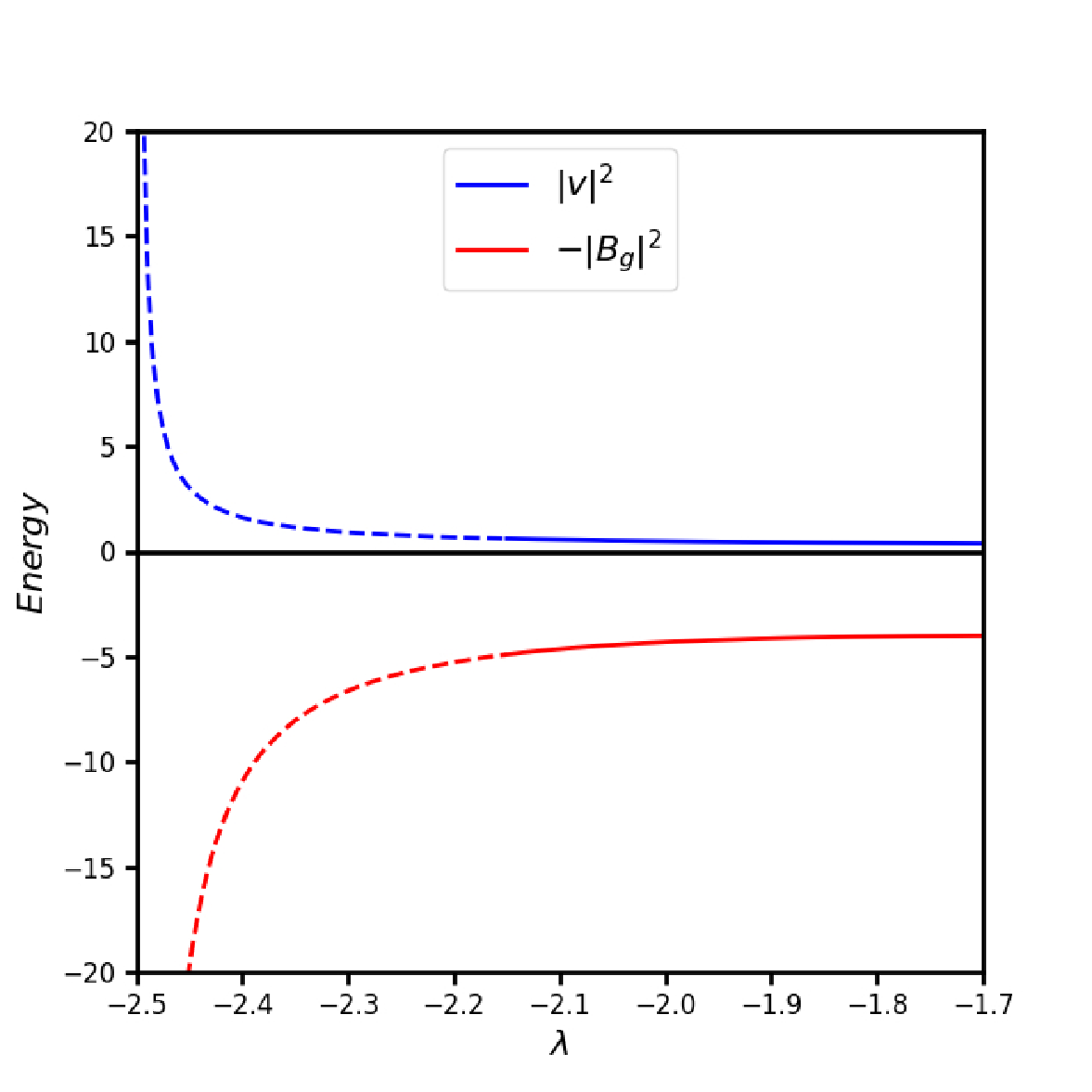}
\caption{Plots for flow and gravito-magnetic energies %multiplied by $\kappa 10^2$
vs $\lambda $ for the scenario leading to final Single Beltrami (macro-scale)
state (see parameters in Fig.5) - macro-scale flow formation scenario
following the increase of inverse-length scale $|\lambda |$).}
\label{Fig.6}
\end{center}
\end{figure}

All of the above scenarios  dealt with (initially) gravito-magnetically
dominated state with a negative total energy.  In these significant
self-gravity states defining length scales were vastly separated.
In each, the structure at one of the scales
vanishes and the DB state relaxes to an SB one; depending on the initial
parameter ranges the final state represents the flows generated at
macro/micro scales.

There is a very interesting corollary to  the second scenario when the
resulting state has microscopic scale flows. Such turbulent
(we will lump short scale and turbulent flows together) flows
have been deeply investigated \citep{msms,lingam,RD_deg,KS_2TRD}
as a source of generating macro-scale flows and magnetic fields
by what has called the Unified Dynamo/ ReverseDynamo mechanism (D-RD).
Thus the catastrophic loss of equilibrium may eliminate the macro-flow,
but (D-RD) could regenerate it; the structure of the new macro-flow
will, surely, be very different from the original. In fact, in the
companion paper, we will investigate this path.

Getting back to the subject matter of this study, one can,
alternatively, start from a positive (total) energy initial
state in which the flow-energy is dominant and self-gravity is
insignificant. Then, even though we have a DB state initially it is
difficult to imagine that the system follows the (i) scenario
while relaxation.

In fact, numerous simulation runs (not displayed
here) show that with such insignificant gravity,
the macro scale flows may form/accelerate only when
the scales are not vastly separated (the condition $\kappa \ll b^2$ is
not satisfied; the reader may consult with the
root-coalescence scenario for plasma case explored for astrophysical
conditions and well studied in \citep{osym,osym2,Kagan,SSMD}).
It was found that the micro-scale energy completely vanishes and
the macro-scale flow formation takes place;
while relaxation $|\lambda |$ is decreasing implying the destroy
of the structure.
Such conditions can be satisfied in the objects with low vortical
character ($b \sim \kappa \ll 1$) or with initially hot
vortical/turbulent rotating flows ($b \sim \kappa < 1$).

\bigskip

We emphasize here that the transformation characteristics
(the rates and direction / mix of energies) are dictated by the initial
conditions of the system. A system may be more or less prone to
a catastrophe depending on its mode of preparation. It must be noted, however,
that the theory presented in this paper can merely work out conditions
(and other critical parameters) for the occurrence of the phenomenon,
but it has no ability to deal with the phenomenon itself. It is clear
that when the system approaches the critical point, the latter will
require a fast scale dynamical description that must include,
inter alia, the transport processes.

\section{Summary and Conclusions}

In present study, we explored the self-gravity driven flow generation
in slowly rotating neutral fluid/gas
found in various astrophysical settings. The self-gravitating system,
obeying what we call Gravito-magnetic field equations (closely resembling
the Maxwell-Lorenz system of electrodynamics), follow a trajectory
similar to that of its electromagnetic counterpart, and can be analyzed
in terms of the Beltrami-Bernoulli equilibrium \citep{osym,osym2}
whose principal part is the Double Beltrami equation that the velocity
field obeys. Under appropriate conditions (determined by the initial
conditions of the system), this DB equilibria can be catastrophically
lost; the resulting Single Beltrami equilibrium can have a much larger
flow kinetic energy than the original system. Such processes can have
significant effects for the formation of macro-scale flows / structures
in galaxies, accretion discs around the massive compact object; for
the dynamics and stability of a rotating star / its atmosphere etc
\citep{Tobias,yso,jet-photon,Chimnoy}.

The process of flow generation manifests in several steps: on a slow
change in the ambient parameters, the DB gravito-magnetofluid state,
becomes unstable, in particular, one of (macro- or micro- scale) the two
Beltrami components tends to disappear and the resulting readjustment
takes place through a catastrophic rearrangement of the energy mix.
In the problem of interest, gravitational energy ends up powering strong
flows. The critical point at which the transformation takes
place depends on the invariants of the collisionless gravito-magnetofluid
dynamics. Self-gravity provides the energy for flow generation. In detail:

\begin{itemize}
\item we have derived the condition of the catastrophic
transformation of energies.  At the critical point, the Double Beltrami
state is reduced to a single Beltrami one. The energy for
macro- and micro-scale velocity generation comes from the gravito-magnetic fields.

\item The most important qualitative result is that well defined
initial conditions can lead to a kinetically rich final
state - in fact all of the gravito-magnetic field energy is converted
to velocity field energy at the catastrophe.
\end{itemize}

In the end, it must be emphasized that our minimal idealized model
analysis is limited to figuring out
what DB states are prone to a catastrophic change, even predicting the critical
parameters. But when the critical stage is reached, we have to look for a totally
time dependent dynamic theory to work out the fast stages of change. The Unified
Reverse Dynamo/Dynamo mechanism \cite{msms,RD_deg,KS_2TRD} could, for instance,
be the basis of  a theory. Such a study will constitute the scope of future work.

YSOs, HH objects, Protoplanetary disks, rotating planets represent
realistic astrophysical conditions for which the theory, developed here, should
be applicable (see e.g., \cite{yso}) and references therein). Having demonstrated
the important role of self-gravity (in an idealized setting) as an important
factor in determining the equilibrium state, we plan to take the theory to the
aforementioned realistic systems.

%%%%%%%%%%%%%%%%%%%%%%%%%%%%%%%

\section{Acknowledgements}

Present work was partially supported by Shota Rustaveli
Georgian National Foundation Grant Project No. FR-22- 8273.
SMM's research is supported by  U.S. DOE under Grant Nos. DE-
FG02-04ER54742 and DE-AC02-09CH11466.

\section{Author contribution statement}
All authors contributed to the theoretical model, analysis,
calculations and reviewed, edited the manuscript.
N.L.S. and S.M.M. contributed to the methodology
and concept of the analytical approach.
L.G., G.S. contributed to the visualization the results preparing the Figures.
All authors contributed to the writing the main text of the
original manuscript

%\bibliography{<yourbibfile>}

\end{document}